\documentclass[aps,prl,twocolumn, twoside,floatfix]{revtex4-1}
\usepackage[dvips]{color}
\usepackage{graphicx}
\usepackage{newlfont}
\usepackage{ulem}

\begin{document}
\title{Ion-trap electrode preparation with Ne$^+$ bombardment}
\author{K. S. McKay}
\author{D. A. Hite}
\author{Y. Colombe}
\altaffiliation{Present Address: Institut f\"ur Experimentalphysik, Universit\"at Innsbruck, Technikerstr. 25, A-6020 Innsbruck, Austria}
\author{R. J\"ordens}
\author{A. C. Wilson}
\author{D.~H.~Slichter}
\author{D. T. C. Allcock}
\author{D. Leibfried}
\author{D. J. Wineland}
\author{D. P. Pappas}

\affiliation{National Institute of Standards and Technology, 325 Broadway, Boulder, Colorado  80305}

\begin{abstract}
 We describe an ex-situ surface-cleaning procedure that is shown to reduce motional heating from ion-trap electrodes.  This precleaning treatment, to be implemented immediately before the final assembly and vacuum processing of ion traps, removes surface contaminants remaining after the electrode-fabrication process.  We incorporate a multi-angle ion-bombardment treatment intended to clean the electrode surfaces and interelectrode gaps of microfabricated traps.  This procedure helps to minimize redeposition in the gaps between electrodes that can cause electrical shorts.  We report heating rates in a stylus-type ion trap prepared in this way that are lower by one order of magnitude compared to a similar untreated stylus-type trap using the same experimental setup.
\end{abstract}

\maketitle

Electric-field noise from ion-trap surfaces can cause motional heating and limit the coherence time of quantum operations to less than that required for fault-tolerant quantum information processing \cite{faulttolerant}.  Heating rates tend to decrease with increasing ion-electrode distance \textit{d}, however smaller ion traps are desired for scalability and faster quantum gates.  The motional heating rate $\dot{\overline{n}}$, defined as the time rate of change of the average motional-state occupation number, is related to the spectral density of the electric-field noise $S_E$ by
 \begin{equation}
S_E(\omega) = {4m \hbar \omega \over q^2} \dot{\overline{n}}(\omega),
\label{eq:spectral noise}
\end{equation}
where $\omega/2\pi$ is the motional frequency of the ion in the trap, \textit{m} is the ion's mass, \textit{q} is its charge, and $\hbar$ is Planck's constant divided by 2$\pi$ \cite{Turchette00}.  The electric-field noise that causes motional heating originates from thermally driven processes \cite{Deslauriers06, Labaziewicz08a}, and can be significantly reduced by operating at cryogenic temperatures.  For example, relative to room temperature, reductions in electric-field noise of 22 dB \cite{Brown11, Hite12}, 19 dB \cite{Chiaverini14}, and 34 dB \cite{Labaziewicz08b} have been reported, corresponding to cryogenic-electrode heating rates as low as 70 quanta/s ($^9$Be$^+$, $\omega/2\pi$ = 2.3 MHz), 6 quanta/s ($^{88}$Sr$^+$, 1.3 MHz), and 2 quanta/s ($^{88}$Sr$^+$, 1.0 MHz), at ion-electrode distances of 40, 50, and 75 $\mu$m, respectively.  In addition, electric-field noise from room-temperature surface-electrode traps has been reduced by approximately 22 dB following in-situ treatments of ion bombardment, corresponding to heating rates as low as 43 quanta/s ($^9$Be$^+$, 3.6 MHz, \textit{d} = 40  $\mu$m) \cite{Hite12}, and 3.8 quanta/s ($^{40}$Ca$^+$, 1.0 MHz, \textit{d} = 100 $\mu$m) \cite{Daniilidis13}.  For comparison, heating rate data and other experimental parameters for some traps are summarized in Table 1, and the corresponding normalized electric-field noise spectral densities $\omega S_E$ are plotted in Fig. 1 as a function of \textit{d} \cite{Hite12}.

\begin{table*}%
\caption{Heating rates $\dot{\overline{n}}$ from electric-field noise in some microfabricated ion traps of various sizes. Electrode treatments and experimental parameters are listed to compare the ``frequency-normalized'' electric-field noise spectral densities $\omega S_E$ for various ion-electrode distances \textit{d} and electrode temperatures.  The data are grouped by electrode temperature and the following electrode treatments: untreated (\textit{UT}), apart from typical microfabrication cleaning with solvents, in-situ treated (\textit{IT}), and pretreated (\textit{PT}), described in this work.}
\begin{ruledtabular}
\begin{tabular}{ccccccccc}
  && $\dot{\overline{n}}$ &\textit{d} &$\omega/2\pi$ &&Electrode&$\omega S_E$&\\
  &Treatment & (s$^{-1}$)&($\mu$m)&(MHz)&Ion&Material&(10$^{-6}$ V$^2$/m$^2$)&Reference\\
\hline
\hline
  1&\textit{UT} &600&40&4.6&$^{25}$Mg$^+$&Au&340&\cite{Epstein07}\\
  2&(300 K) & 500 & 50 & 1.3 &$^{88}$Sr$^+$&Au& 83  &\cite{Chiaverini14}\\
  3&& 410 & 63 & 1.4 &$^{40}$Ca$^+$&Al& 33  &\cite{Doret12}\\
\hline
  4&\textit{UT} & 70 & 40 & 2.3 &$^{9}$Be$^+$&Au& 3.6  &\cite{Brown11}\\
  5&(5 K) & 6 & 50 & 1.3 &$^{88}$Sr$^+$&Au& 1.0  &\cite{Chiaverini14}\\
  6&& 2.1 & 100 & 0.8 &$^{88}$Sr$^+$&Nb& 0.025  &\cite{Wang10}\\
\hline
    7&\textit{IT}&43&40&3.6&$^{9}$Be$^+$&Au&5.4&\cite{Hite12}\\
  8&(300 K)&3.8&100&1.0&$^{40}$Ca$^+$&Cu/Al&0.15&\cite{Daniilidis13}\\
\hline
    9&\textit{UT} (300 K)&16,000&40&3.6&$^{9}$Be$^+$&Au&2100&\cite{Hite13}\\
    10&\textit{PT} (300 K)&930&40&3.6&$^{9}$Be$^+$&Au&120&\cite{Hite13}\\
\hline
  11&\textit{UT} (300 K)&387&62&4.0&$^{25}$Mg$^+$&Au&170&\cite{Arrington13}\\
  12&\textit{PT} (300 K)&30&60&4.4&$^{25}$Mg$^+$&Au&14&This Work\\
\end{tabular}
\end{ruledtabular}
\end{table*}

\begin{figure}[b]

           \includegraphics[width=.45\textwidth]{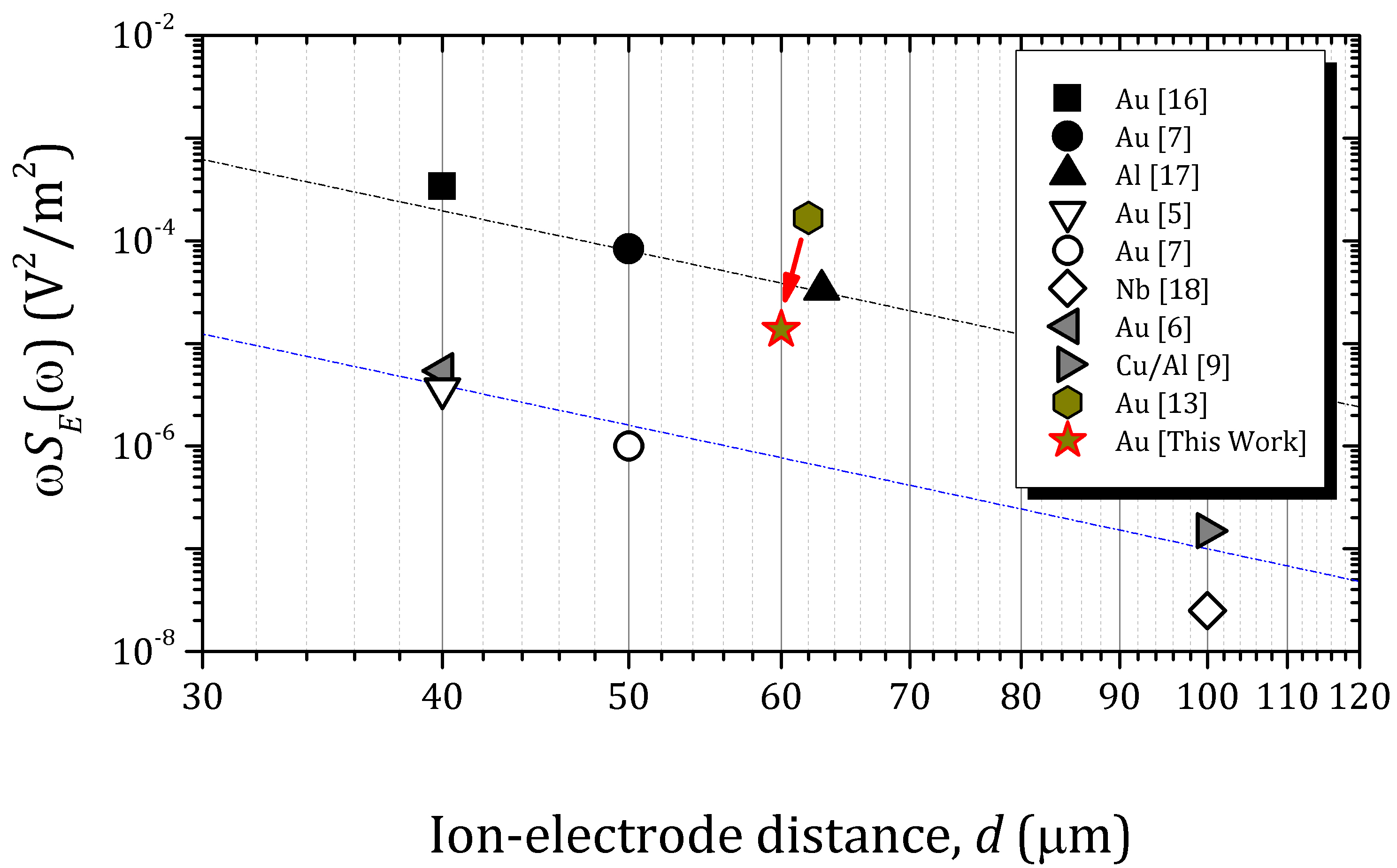}

    \caption{Normalized electric-field noise spectral densities in selected low-noise microfabricated traps of various sizes and electrode materials.  Since $S_E$ exhibits an approximate $1/\omega$ dependence in many ion-trap experiments, we normalize data taken at different frequencies $\omega/2\pi$ onto a common curve by plotting $\omega S_E$.  Dotted lines indicating a 1/$\textit{d}^{4}$ distance scaling are plotted as guides to the eye.  Open symbols indicate data from cryogenic traps with electrodes cooled to liquid helium temperatures.  Electrode materials are indicated in the legend.}

\end{figure}

Despite this progress in reducing motional heating rates through cryogenic operation or in-situ treatments, electric-field noise in these traps typically remains well above estimates for thermal voltage noise (Johnson noise) from the electrode circuitry, and therefore is commonly referred to as ``anomalous heating''.  While this anomalous heating is thought to be related to surface contamination on the trap electrodes, a deeper understanding of its physical origin will help to solve this problem.  In an experiment that tested the effects of contamination from air exposure and vacuum baking on a surface-electrode trap that was previously treated with in-situ Ar$^+$ bombardment \cite{Hite12}, heating rates were measured to increase from 50 to 930 quanta/s ($^9$Be$^+$, 3.6 MHz, \textit{d} = 40  $\mu$m), after a 48-h exposure to air followed by vacuum baking at 180 $^\circ$C for approximately 100 h (see Fig. 5(c) in Ref.~\cite{Hite13}).  This corresponds to a 12 dB reduction in the electric-field noise spectral density as compared to that measured before the Ar$^+$-cleaning treatments (Table I, \# 9 and 10).  These data indicate a potential benefit resulting from a precleaning treatment prior to trap assembly.  While it is believed that in-situ treatments are required for full removal of the dominant source of this anomalous heating \cite{Hite13}, it would be advantageous to implement an ex-situ precleaning treatment to reduce heating rates, thereby enabling new experiments, operating at either room or cryogenic temperatures, without the need for in-situ cleaning capabilities.  It may be that in-situ cleaning treatments can be combined with cryogenic cooling to reduce heating rates further; however, implementing in-situ ion-bombardment treatments in a cryogenic ion-trap setup is challenging, given the necessity to provide optical access, and to incorporate cryogenic, vacuum, and gas-handling components.

Ion bombardment tends to redeposit sputtered metal into the interelectrode gaps.  This redeposited metal can cause electrical shorts in regions of the gaps that are shielded by the electrodes from ion bombardment.  Since the etch rate from sputtering is observed to be greater than the rate of redeposition, designing the trap-electrode and ion-beam geometry to eliminate the shielding of the gaps can mitigate this problem.  One could also optimize the dose of the in-situ treatment required to achieve the desired heating rate by employing treatment parameters (i.e., beam energy, mass, and incident angle) that minimize the sputter yield of the electrode material and concurrently maximize that of the surface contaminants \cite{Eckstein07}.  In addition, one could perform an ex-situ precleaning treatment just prior to final trap assembly, allowing ion bombardment from multiple angles to remove any metal redeposited in the gaps.  There is evidence from Ref.~\cite{Hite12} that higher sputtering energies may be necessary to achieve very low heating rates at room temperature. This suggests a requirement for relatively high-energy surface modifications, e.g., removal of constituents with relatively low sputter yield \cite{footnote1}.  However, higher sputtering energies likely exacerbate the redeposition problem.  In addition to the potential benefit to traps without in-situ cleaning capabilities, ex-situ precleaning treatments may also help to reduce this redeposition problem by allowing a gentler in-situ treatment.

In this work, we used a multi-angle precleaning procedure using high sputter doses to remove contaminants while simultaneously removing redeposited metal from the gaps between electrodes.  We compare heating rates in two similar (but not identical) stylus-type ion traps, one with and one without the precleaning treatment (Table I, \# 12 and 11, resp.), and find a heating rate in the precleaned trap lower by an order of magnitude.  Making use of in-situ Auger electron spectroscopy, we correlate this reduction in electric-field noise with the reduction in surface contaminants due to the precleaning procedure.

Using a microfabricated stylus trap, described in detail in Ref.~\cite{Arrington13} (Table I, \# 11), we reported an untreated trap-electrode heating rate of 387 $\pm$ 15 quanta/s for $d =$ 62 $\mu$m above an 80-$\mu$m diameter Au stylus post (Fig. 1, hexagon symbol).  The corresponding electric-field noise, which is well above estimates for thermal voltage (Johnson) noise, is thought to arise from dynamical processes of contaminants on the surface that cause contact-potential fluctuations.  These contaminants have many potential sources, including residues from the fabrication processing, vacuum baking for ultra-high vacuum (UHV) operation, or other adventitious contamination from air exposure.

In order to test the dependence of ion motional heating on the amount of surface contamination, and the benefits of a precleaning treatment, in this work we prepared two nominally identical traps that were fabricated on the same wafer.  Both traps were precleaned as described next, exposed to air for $\sim$ 24 h, and then vacuum-baked together in the ion-trap chamber.  The precleaning consisted of a series of treatments of Ne$^+$ bombardment at three different angles: 10 min at + 30$^{\circ}$ incident angle (with respect to the normal of the surface, see Fig. 2), 10 min at - 30$^{\circ}$, and 20 min at 0$^{\circ}$, all treatments using 2 kV and 30 $\mu$A/cm$^2$ with a $\sim$ 3-mm focus diameter of the incident ion beam, at a Ne pressure of $\sim$ 6 $\times$ 10$^{-3}$ Pa \cite{footnote2}.  The initial angled treatments are intended to remove most of the contaminants away from the surface and minimize any embedded contaminants. The final normal-incidence treatment cleans shadowed redeposition of metal (and other deleterious impurities) from the gaps in the electrodes.  One of the traps, shown in Fig. 2, was used to trap ions and compare heating rates with the untreated trap of Ref.~\cite{Arrington13}.  The duplicate trap was mounted to a sample transfer stage in the ion-trap chamber and subsequently transferred, in UHV, to a surface-science chamber for the post-bake analysis.

\begin{figure}[b]

           \includegraphics[width=.37\textwidth]{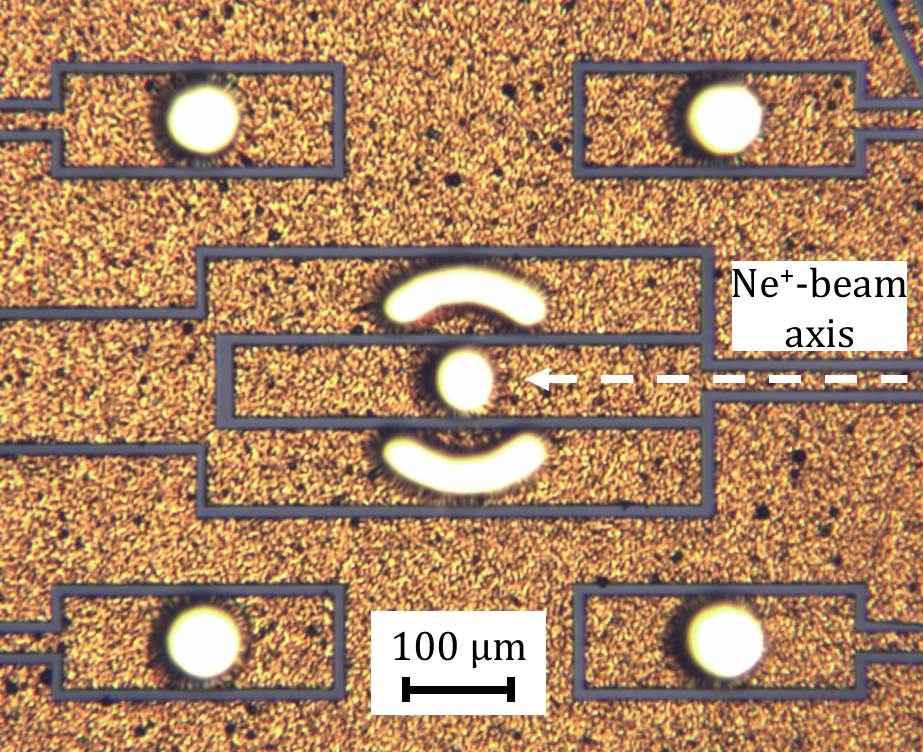}

    \caption{Top-view optical micrograph of the as-fabricated stylus trap used in this work before the precleaning treatment.  The electroplated Au electrodes are formed in two layers, a 10-$\mu$m thick base layer (grainy gold color) and the 200-$\mu$m tall stylus posts (appearing white) protruding from the base layer \cite{Arrington13}.  The interelectrode gaps are 10 $\mu$m wide.  The orientation of the Ne$^+$ sputter beam with respect to the electrodes is indicated.  The precleaning treatment angles are $\pm$ 30$^{\circ}$ and 0$^{\circ}$ with respect to the normal of the surface (out of the page).  $^{25}$Mg$^+$ ions are trapped at a height $d =$ 60 $\mu$m above the 60-$\mu$m diameter center post. Other than the geometric differences on the chip, this trap is electrically the same as the trap in Ref.~\cite{Arrington13}, and uses the same experimental setup for the heating rate measurements.}

\end{figure}

Prior to the precleaning treatment or vacuum baking, Auger electron spectroscopy (AES) of the untreated Au electrodes on the as-fabricated duplicate trap chip indicates the presence of carbonaceous contamination (1 - 2 carbon monolayers (ML) thick), previously correlated with anomalous heating \cite{Hite12}, along with small amounts of other contaminants (S, O, Fe) from the electroplating bath used for gold deposition (Fig. 3-a).  Immediately after applying the Ne$^+$ precleaning treatment to the duplicate trap-electrode surfaces, Auger spectra collected in situ signify a clean Au surface free of these contaminants (Fig. 3-b).  Before vacuum baking, air exposure of these precleaned surfaces for $\sim$ 24 h typically results in a reacquiring of less than 0.5 ML of oxygen-free adventitious carbon.  The level of contamination on the actual ion-trap electrodes was also monitored using AES during the steps of the precleaning treatment to confirm comparable results.  Upon vacuum baking for UHV processing of the assembled ion-trap chamber, the thickness of the carbonaceous layer grows slightly, presumably from hydrocarbons desorbing from the vacuum-chamber walls and other surfaces.  Nevertheless, vacuum baking of these precleaned and air-exposed surfaces typically results in less surface contamination than the untreated, as-fabricated electrodes.  The approximate carbon coverage on the precleaned, air-exposed, then vacuum-baked duplicate trap was determined from in-situ AES to be $\sim$ 0.6 ML (Fig. 3-c).  Since the actual trap chip and the duplicate trap chip were exposed to air and vacuum baked together in the assembled ion-trap chamber, we assume that the actual trap chip has the same level of contamination as the duplicate trap.

\begin{figure}[t]

           \includegraphics[width=.42\textwidth]{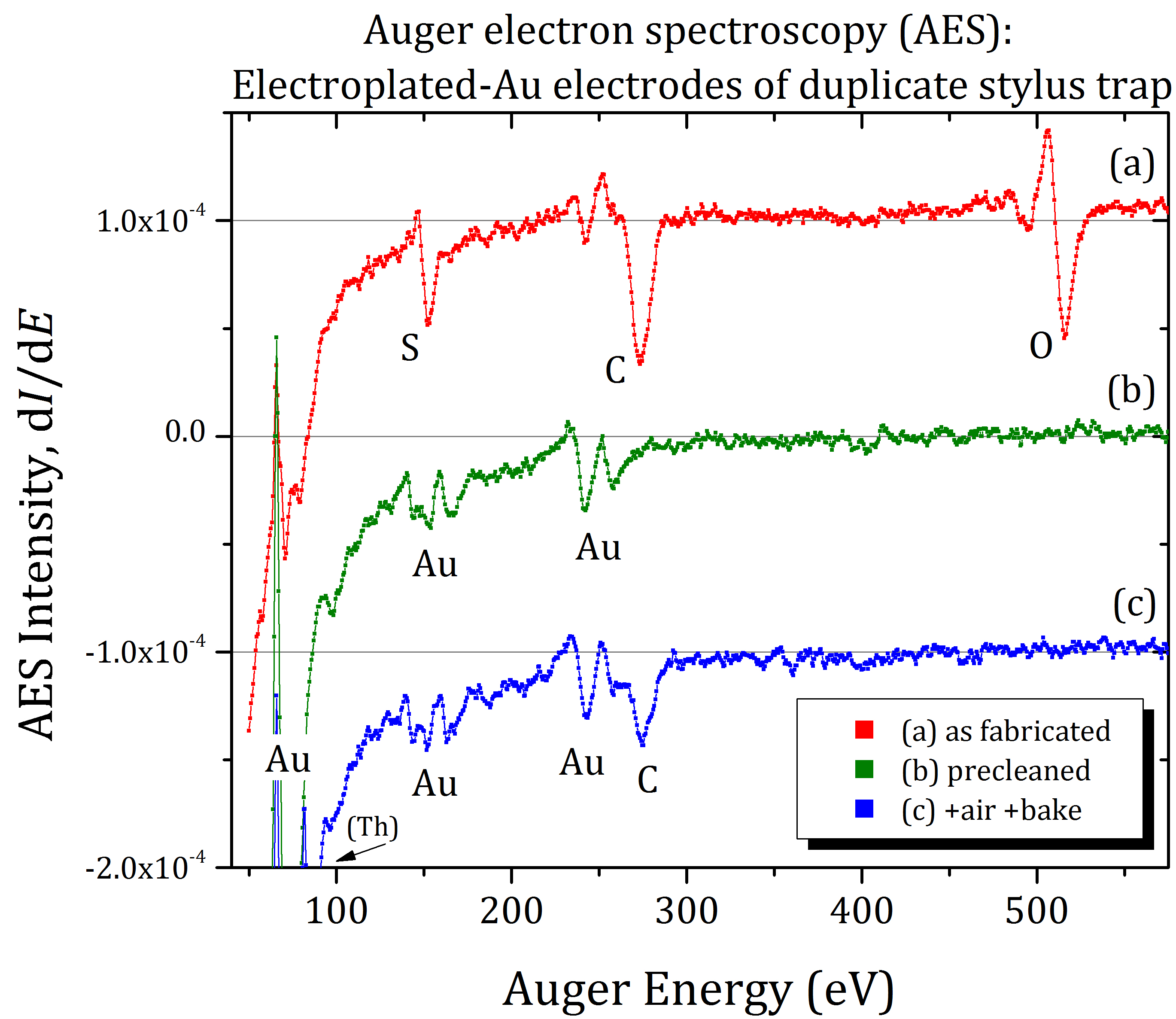}

    \caption{Surface analysis of the electroplated-Au electrodes on the duplicate trap chip, (a) before and (b) immediately after the precleaning treatment.  This companion sample was then exposed to identical environments as the actual trap chip used to measure heating rates, namely air exposure for 24 h and vacuum baking at 180 $^{\circ}$C for approximately 100 h (together in the same chamber).  Spectrum (c) was acquired from the duplicate trap chip in situ, after operating the Mg source used in the ion-trap chamber and then UHV transfer to the surface analysis chamber.  No Mg was detected on the duplicate trap chip. We note the presence of a signal from thallium, which is driven to the surface from the bulk material upon baking of the electroplated, soft-gold electrodes.  Traces are offset vertically for clarity.}

\end{figure}

The lowest motional heating rate of the precleaned ion trap was measured to be 30 $\pm$  2 quanta/s (Table I, \# 12), for $d =$ 60 $\mu$m above a 60-$\mu$m diameter Au stylus post (Fig. 1, star symbol).  This corresponds to an 11 dB improvement in electric-field noise compared to the untreated stylus trap of Ref.~\cite{Arrington13} (Table I, \# 11).  Moreover, this is consistent with the level of improvement (12 dB) observed in the air-exposure/vacuum-baking test of Ref.~\cite{Hite13} (Table I, \# 9 and 10), discussed above.  Therefore, this suggests that the significantly lower heating rate reported here is due to surface modifications made during the precleaning treatment, and that the electric-field noise from surfaces sensed by trapped ions is directly related to the coverage of adsorbates on the surface.

There are slight differences between the untreated trap of Ref.~\cite{Arrington13} and the precleaned trap, whose heating rates are being compared in this study.  Most of these differences are geometric design modifications made to the stylus posts, primarily to minimize shadowing of the gaps near the posts during sputter treatments that can cause electrical shorting.  Beside increasing the height of the stylus posts to 200 $\mu$m above the chip (compared to the 150-$\mu$m posts used in Ref.~\cite{Arrington13}), the diameter of the center post was reduced to 60 $\mu$m (Fig. 2), from the Ref.~\cite{Arrington13} design of 80 $\mu$m.  Due to these geometric changes, simulations of the trap indicate that the ion is trapped approximately 60 $\mu$m above the center post, as compared to 62 $\mu$m in the trap in Ref.~\cite{Arrington13}.  The different geometries of the untreated and precleaned traps should correspond to slightly different electric-field noise spectral densities at the ion for a given voltage noise on the electrode surfaces.  Assuming either a 1/$\textit{d}^{4}$ or 1/$\textit{d}^{2}$ distance scaling for the noise, we estimate that the precleaned trap should have 10$\%$ or 15$\%$ less noise, respectively, due to geometrical considerations.  Therefore, we conclude that these geometric differences cannot account for the observed reduction in heating rate.

We checked that the heating rate in the precleaned trap was not limited by extraneous technical noise.  This can arise from voltage noise from the computer-controlled digital-to-analog converters and amplifiers that apply static potentials to the trap.  Therefore, to exclude this possibility, we applied static (dc) potentials generated by batteries.  We also checked for ion state de-pumping during the wait period in the heating rate measurements by using a fast shutter to eliminate possible stray light from the acousto-optical switches. During these tests, we observed no measurable differences in the heating rate.

Preliminary tests of the cleaning procedure on other electrode materials indicate that, in addition to gold, Cu and Al electrodes may also benefit from such a precleaning treatment. However, for Nb electrodes, our Auger analysis indicates the presence of a very low-sputter-yield refractory-metal carbide following the precleaning treatment.  Niobium carbide forms when the native oxide of Nb is covered with adventitious carbon and undergoes ion bombardment \cite{Grundner80}. This may preclude the use of ion-bombardment precleaning treatments on Nb electrodes.

This research was funded by the Office of the Director of National Intelligence (ODNI), Intelligence Advanced Research Projects Activity (IARPA). All statements of fact, opinion, or conclusions contained herein are those of the authors and should not be construed as representing the official views or policies of IARPA or the ODNI.  We also acknowledge the support of ONR and the NIST Quantum Information Program.  We thank the Sandia National Laboratories' Metal Micromachining Program and CAMD for fabricating the stylus traps used in this work.  This article is a contribution of the U.S. Government, and is not subject to U.S. copyright.


\begin{references}

\bibitem{faulttolerant} J. Preskill, Proc. R. Soc. London, Ser. A \textbf{454}, 385 (1998); A. S$\o$rensen and K. M$\o$lmer, Phys. Rev. A \textbf{62}, 022311 (2000); E. Knill, Nature \textbf{463}, 441 (2010).

\bibitem{Turchette00} Q. A. Turchette, \textit{et al.}, Phys. Rev. A \textbf{61}, 063418 (2000).

\bibitem{Deslauriers06} L. Deslauriers, S. Olmschenk, D. Stick, W. K. Hensinger, J. Sterk, and C. Monroe, Phys. Rev. Lett. \textbf{97}, 103007 (2006).

\bibitem{Labaziewicz08a} J. Labaziewicz, Y. Ge, P. Antohi, D. Leibrandt, K. R. Brown, and I. L. Chuang, Phys. Rev. Lett. \textbf{100}, 013001 (2008).

\bibitem{Brown11} K. R. Brown, C. Ospelkaus, Y. Colombe, A. C. Wilson, D. Leibfried and D. J. Wineland, Nature \textbf{471}, 196 (2011).

\bibitem{Hite12} D. A. Hite, \textit{et al.}, Phys. Rev. Lett. \textbf{109}, 103001 (2012).

\bibitem{Chiaverini14} J. Chiaverini and J. M. Sage, Phys. Rev. A \textbf{89}, 012318 (2014).

\bibitem{Labaziewicz08b} J. Labaziewicz, Y. Ge, D. R. Leibrandt, S. X. Wang, R. Shewmon, and I. L. Chuang, Phys. Rev. Lett. \textbf{101}, 180602 (2008).

\bibitem{Daniilidis13} N. Daniilidis, S. Gerber, G. Bolloten, M. Ramm, A. Ransford, E. Ulin-Avila, I. Talukdar, and H. H\"{a}ffner, ArXiv 1307.7194v1 (2013).

\bibitem{Hite13} D. A. Hite, Y. Colombe, A. C. Wilson,D. T. C. Allcock, D. Leibfried, D. J. Wineland, and D. P. Pappas, MRS Bull. \textbf{38}, 826 (2013).

\bibitem{Eckstein07} W. Eckstein, in \textit{Sputtering by Particle Bombardment}, Topics in Applied Physics Vol. 110, edited by R. Behrisch and W. Eckstein (Springer, Berlin, 2007), p.33.

\bibitem{footnote1} Generally, use of Ar$^+$ results in higher sputter yields, however the lower sputter yields for Au using Ne$^+$ are desired to mitigate redeposition. See Ref.~\cite{Eckstein07}.

\bibitem{Arrington13} C. L. Arrington, \textit{et al.}, Rev. Sci. Instrum. \textbf{84}, 085001 (2013).

\bibitem{footnote2} Base pressures should be low enough that the adsorption rate of residual gas species on the surface is less than the sputter rate of the impinging ion beam, e.g., to avoid oxide or nitride formation \cite{Eckstein07}.

\bibitem{Grundner80} M. Grundner and J. Halbritter, J. Appl. Phys. \textbf{51}, 397 (1980).

\bibitem{Epstein07} R. J. Epstein, \textit{et al.}, Phys. Rev. A \textbf{76}, 033411 (2007).

\bibitem{Doret12} S. C. Doret \textit{et al.}, New J. Phys. \textbf{14}, 073012 (2012).

\bibitem{Wang10} S. X. Wang, Y. F. Ge, J. Labaziewicz, E. Dauler, K. Berggren, I. L. Chuang, Appl. Phys. Lett. \textbf{97}, 244102 (2010).

\end{references}
\end{document}